\documentstyle[aps,epsfig,preprint]{revtex}
\begin{document}

\title{High-Current Field Emission from an Atomic Quantum Wire}

\author{A. Lorenzoni$^{1,2}$, H.E. Roman$^{3}$,
	F. Alasia$^{2}$ and R.A. Broglia$^{1,2,4}$}

\address{$^1$Dipartimento di Fisica, Universit\`a di Milano,
Via Celoria 16, I-20133 Milano, Italy.}

\address{$^2$INFN, Sezione di Milano, Via Celoria 16, I-20133 Milano, Italy.}

\address{$^3$Institut f\"ur Theoretische Physik III, Universit\"at Giessen,
Heinrich-Buff-Ring 16, 35392 Giessen, Germany.}

\address{$^4$The Niels Bohr Institute, University of Copenhagen,
2100 Copenhagen, Denmark.}

\date{\today}

\maketitle

\smallskip

\begin{abstract}
Linear chains of carbon atoms have been proposed  \cite{Smalley1}
as the electron emitting
structures of open tip carbon nanotubes subject to an electric field.
To better understand the implications of the results of Smalley and 
collaborators, the electromagnetic response of linear carbon chains
to both static and dynamics fields have 
been studied, making use of ab-initio methods.
It is found that the associated emission currents, plotted as a function of the 
bias potential, follow Fowler-Nordheim intensity-voltage curves typical
of the field emission of metallic tips. Under standard bias conditions, 
linear carbon chains of one nanometer of length are expected to
deliver currents of the order of one microamp\`ere.
These systems behave, furthermore, as conducting needles in photoabsorption
processes. Linear carbon chains are thus likely to constitute the ultimate 
atomic-scale realization of metallic wires.
\end{abstract}

\newpage
In studying the field emission of electrons from individually mounted 
carbon nanotubes, Smalley and collaborators have found \cite{Smalley1}  
a conspicuous increase of the yield when the nano-tube tips were opened
by laser evaporation, an increase which became accentuated when the open
tips cooled down to room temperature. This behaviour lead the authors
to conclude that the emitting structure was an "atomic wire" of
10 to 100 sp-bonded carbon atoms pulled from the open graphene sheet of
the nanotube by the electric field.

Triggered by these remarkable results, we have studied the quantal 
properties of $C_n$ chains ($n=3-11$), making use of ab-initio 
methods. We have found that when the linear chains of carbon atoms , 
which display an electronic structure very close to the cumulenic
form ($\cdots\!=\!C\!=\!C\!=\!C\!=\!C\!:$) , are subject to a bias 
field of intensity 
$0.6-0.7\; V/\AA$, currents of the order of $1\;\mu A$ 
are obtained , in overall agreement with the experimental findings.
Even-n chains lead to currents which are a factor $\approx 10^2$
more intense than those associated with the field emission of 
electrons from odd-n chains. Shining photons on the $C_n$ chains
leads to a frequency-dependent photoabsorption cross section which 
essentially coincides with that of a classical metallic needle 
of the same shape. These results lend strong support to the
conjecture \cite{Smalley1}
that linear chains of carbon atoms can be viewed
as atomic-scale metallic wires.   
The electronic structure calculations of the $C_n$ chains have 
been carried out in the local density approximation (LDA) including
exchange-correlation effects according to the parametrization of
Perdew and Zunger \cite{Perdew}, while the role of the carbon atoms were
taken into account in terms of norm-conserving pseudopotentails
(\cite{Troullier}, cf. also ref. \cite{Broglia1}). 
The resulting bond length is constant and equal to
$1.31\;\AA$ for n-odd chains, while it alternates by less than
$2\%$ around this value in the case of n-even chains 
(cf. also ref. \cite{Ragh}).The constancy of the bond length can be seen 
from Fig.1(A), where the electronic density of the linear chain $C_8$ is 
displayed. It is also found an almost complete screening of the field 
along the chain, as it is the case for perfect conductors
(Fig.1(B), cf. also \cite{Smalley2}).

The electron emission characteristic were calculated in the WKB
approximation making use of the LDA potential felt by the electron
\cite{Gomer}. Typical current versus voltage (I-V) curves are shown 
in Fig. 2. They can be compared with the Fowler-Nordheim equation for the 
field emission $I=a F^2 \exp(-b/F)$, where $a$ and $b$ are costants that
depend on the electronic work-function of the system as well as
on the image correction term, a quantity which depends weakly on
the electric field strength at the emitting surface $F=V/kR_0$.
Here $R_0$ is the radius of the tip and $k$ is a constant of the order
of 10 \cite{Collins}.
Making use of the fact that the linear chains under discussion
have a value $R_0 \approx 1.2 \AA$ and of the results shown in Fig.2,
it is seen that emission currents of the order of $1\;\mu A$ 
are obtained with a bias voltage $V$ of the order of 
$40 V$, in overall agreement with the experimental findings
\cite{Smalley1}. Because the emitted electrons arise essentially from the 
occupied levels lying closest to the Fermi energy (HOMO) \cite{rimando1}, 
the emitted electrons are rather monoenergetic with 
energies that fluctuate less than $5\;\%$ around the mean value of 
$5 eV$.


To gain further insight into the properties of the $C_n$-chains,
we have subject them to a time dependent electromagnetic field.
The associated longitudinal photoabsorption cross section calculated
in the time-dependent LDA (cf. ref. \cite{Broglia1}), essentially displays a 
single peak which collects $\approx 90\%$ of the oscillator strength.
The corresponding energy centroids $\hbar \omega(C_n)$ are
displayed in Fig.3 \cite{rimando2}. Theory provides an excellent account of the 
experimental findings \cite{Forney} (cf. also ref.\cite{Berch}).

The values of $\hbar \omega(C_n)$ shown in Fig.3 can be compared with the 
results of the relation  
$\hbar \omega_s=\sqrt{L} \hbar \omega_0$ (cf. refs. \cite{Broglia2} and 
\cite{Berch}).
This expression is a generalization
of the Mie resonance expression, and describes the surface plasmon 
of elongated metallic particles, in terms of the bulk plasmon frequency
$\omega_0$ \cite{Bohren}. The quantity
$L=\frac{1-e^2}{e^2}(-1+\frac{1}{2e}\log(\frac{1+e}{1-e}))$ is the
depolarization factor for vibrations along the symmetry axis of the
system.The quantity $e$ is related to the ratio of short to long axis
$R_{\perp}/R_{\parallel}$ according to
$e^2=1-(R_{\perp}/R_{\parallel})^2$.  The quantity $\hbar \omega_s$ 
provides a very good fit to the centroid 
energies $\hbar \omega(C_n)$ with $\hbar \omega_0=24 eV$ (cf. Fig.3).
This value is quite close to the one obtained inserting, in the standard 
plasmon relation $\omega_0^2=4\pi e^2 n/m$, the density of graphite.

We have also calculated the polarizability of the system, and displayed it
in Fig.4. These results can again be compared with the polarizability
associated with a perfect conductor of the same dimension, that is
$
  \alpha_s=v[(\epsilon-1)/(1+L(\epsilon-1))]
$
(cf.ref.\cite{Bohren}).  
In this expression $v$ and $\epsilon$ are the volume and the dielectric
constant of the system, respectively. The long wavelength limit of the 
time dependent LDA results are accurately
reproduced by the function $\epsilon=0.34N^{1.24}+1$, which already for a 
$10\;nm$ linear chain leads to a dielectric constant of the order of
$10^2$, indicating the conducting properties of the system.
 
Linear carbon chains seem to pass with ease, some of the most obvious and
stringent tests needed to qualify as metallic 
atomic wires. In particular, chains
with an even number of atoms behave as prolific emitters of monoenergetic
electrons as well as sensitive antennas.

Financial support by NATO under grant CRG 940231 is gratefully acknowledged. 
We are also indebted for financial support to INFM Advanced Research
Project Class.




\begin{thebibliography}{99}
\bibitem{rimando1}
In keeping with this discussion, it is noted that for
one-dimensional systems like 
the ones under discussion, quantal size effects (QSE) are of importance, 
and the total current emitted under the influence of an external electric
field is the summed contribution of each single-particle quantum 
level. Thus, the equivalent to the Fowler-Nordheim (FN) equation for
field emission from a carbon chain,taking into account QSE is 
$I=a'F \exp(b'/F)$, that is ,a relation similar to the FN-equation
but with a prefactor linear in the electric field.
While the standard FN-equation provides a 
reasonable fit to the numerical results displayed in Fig.2, 
the results of the expression linear in F essentially coincides with 
the WKB results.
\bibitem{rimando2}
To be noted is that even-n carbon chains are stable when they are closed
[11,12]. Under the influence of an external field these chains can be 
assumed to be linear, and the plasmon frequency essentially coincides with
that of the neighbouring odd-n Carbon chains.
\end{thebibliography}
\end{document}